\UseRawInputEncoding
\documentclass[%
 reprint,
 amsmath,amssymb,
 aps,
]{revtex4-2}

\usepackage{graphicx}
\usepackage{dcolumn}
\usepackage{bm}

\begin{document}

\preprint{APS/123-QED}

\title{Wave function propagation in a two-dimensional\\paramagnetic semiconductor from an impurity}

\author{Josh Wanninger}
\author{Gonzalo Ordonez}%
\affiliation{%
 Butler University\\
}%


\date{\today}

\begin{abstract}
We simulated modifications to a model of a two-dimensional paramagnetic semiconductor called the half-BHZ model, also known as the QWZ model, and simulated a modified full BHZ model, where a time reversal pair is introduced. Our modifications to the models include adding single and multiple impurities connected to the lattices or as a connection between the time-reversal pairs. We employed the Julia programming language to show how to speed up calculations for time evolutions. By simulating the time evolutions, we could observe the differences in the effects of these modifications. Our simulations showed the presence of scattering behavior associated with the infinite QWZ model topological states. Moreover, we observed scattering and absorption behavior related to the parameters and placements of impurities and Hamiltonian imaginary component's symmetry or anti-symmetry. These tools and early results lay the foundations for developing electronic devices that use the models' unique scattering and absorption behaviors and explore more complex and physically accurate modifications to the models.
\end{abstract}

\maketitle


\section{Introduction}

Topological insulators are materials that can conduct electrons along their edges, while their bulk acts as an insulator \cite{Hasan10}. 

A remarkable feature of topological insulators is the absence of back-scattering of electrons that move along the edges, for sufficiently wide lattices \cite{Asboth16}. In theory, this allows for perfect conductance along  the edges \cite{Dang15}.  Adding impurities to the structure of the topological insulator can alter the local conductance of these materials, allowing the current to be steered in different directions \cite{Dang15,Winters22}. This can be of great use in electronic components such as transistors, allowing directional control of current with minimal energy loss.

Topological insulators have only been realized at extremely cold temperatures because the traveling electrons require a very small bulk energy gap where the conducting edge states may exist. However, Reis et al \cite{Reis17} have shown that Bismuthene on a SiC substrate has a larger energy gap that may allow for conductance at higher temperatures.  

Here we will numerically investigate the addition of impurities to the material and as a way to create and control edge states. 

As a simple model of topological insulators, we will consider a simple two-dimensional lattice model, the BHZ model. This model itself is built upon the QWZ model \cite{Asboth16}. We will investigate the propagation of a single-electron wave function for finite two-dimensional lattices.

The QWZ and BHZ models are based on a lattice with two internal states for each site to represent the paramagnetic states. Both models allow the wave function probabilities to move into adjacent sites on the right by modifying the internal state and to adjacent sites on the left by the Hermitian of the right operation. The vertical and horizontal wave function scattering have different operations as well.

These models are simulated using their Hamiltonian matrix equations given in their respective subsections below. These matrices represent the two-dimensional lattice onto a two-dimensional matrix. The diagonal of this lattice holds the internal states for each site of the lattice, and the connection can be found above and below the diagonal. The Hamiltonian can grow rapidly from small lattices where a 3x3 lattice requires an 18x18 matrix for the QWZ model and a 36x36 matrix for the modified BHZ model.

The construction of the Hamiltionian for the QWZ and BHZ models in the present work is based on previous work  by D. Winters \cite{Winters22}.

\subsection{The QWZ Model}

The Qi-Wu-Zhang model (QWZ) \cite{Asboth16} is a building block for the BHZ model. The equation of the Hamiltonian is given equation \ref{QWZ}. The Hamiltonian matrix can be easily modified from this base matrix. Our modification includes a time-reversal pair to build a BHZ model and adds impurities to both models. For the QWZ model, a simple impurity was added to the edge of the lattice. This impurity acts the same way as a lattice site but only connects to its single adjacent site. The parameters of this impurity site can also be modified to have different values from the lattice sites. A common impurity modification changes the internal site potential compared to the lattice site potential. 

 \begin{equation}
    \label{QWZ}
     \begin{aligned}
      & \widehat{H}_{\rm QWZ} \\
      & =\sum_{m_x=1}^{N_x-1} \sum_{m_y=1}^{N_y}\left(\left|m_x+1, m_y\right\rangle\left\langle m_x, m_y\right| \otimes \frac{\hat{\sigma}_z+i \hat{\sigma}_x}{2}+h.c.\right) \\
       & +\sum_{m_x=1}^{N_x} \sum_{m_y=1}^{N_y-1}\left(\left|m_x, m_y+1\right\rangle\left\langle m_x, m_y\right| \otimes \frac{\hat{\sigma}_z+i \hat{\sigma}_y}{2}+h.c.\right) \\
       & +u \sum_{m_x=1}^{N_x} \sum_{m_y=1}^{N_y}\left|m_x, m_y\right\rangle\left\langle m_x, m_y\right| \otimes \hat{\sigma}_z
     \end{aligned}
\end{equation}

For an impurity attached to a lattice site with coordinates $(m_{x0}, m_{y0})$, the impurity Hamiltonian is given by Eq. (\ref{impurity}) where $\sigma_w = \sigma_x$ if the impurity is attached to either the left or right edge of the lattices, and $\sigma_w = \sigma_y$ if the impurity is connected to either the top or bottom edge. The total QWZ Hamiltonian including the impurity is then $\widehat{H}_{\rm QWZ} +\widehat{H}_{\rm IMP}$.

 \begin{equation}
     \begin{aligned}
         \label{impurity}
      & \widehat{H}_{\rm IMP} 
        = \epsilon \left|d\right\rangle\left\langle d \right| \otimes \hat{\sigma}_z \\
        &+ \left(\left|d\right\rangle\left\langle m_{x0}, m_{y0}\right| \otimes \frac{\hat{\sigma}_z+i \hat{\sigma}_w}{2}+h.c.\right) 
    \end{aligned}
\end{equation}

\subsection{The BHZ Model}
The modified Bernevig–Hughes–Zhang model (BHZ) \verb+(Source #)+ adds a time reversal pair to the QWZ model shown Eq. (\ref{BHZ}). The original QWZ Hamiltonian $\widehat{H}_{\rm QWZ}$ is associated with an up-spin state of the particle, while the time-reversal pair $\widehat{H}_{\rm QWZ}^*$ is associated with a down-spin state. Including the impurity, the complete Hamiltonian for the modified BHZ is shown in Eq. (\ref{final}).

 \begin{equation}
    \label{BHZ}
    \widehat{H}_{\rm BHZ} =\widehat{H}_{\rm QWZ} \otimes \left|\uparrow\rangle \right. +  \widehat{H}_{\rm QWZ}^* \left|\downarrow\rangle \right.      
\end{equation}

 \begin{equation}    
    \label{final}
    \widehat{H}_{\rm BHZ} = \left( \widehat{H}_{\rm QWZ} +\widehat{H}_{\rm IMP}\right)\otimes \left|\uparrow\rangle \right. +  \left( \widehat{H}_{\rm QWZ} +\widehat{H}_{\rm IMP}\right)^* \otimes \left|\downarrow\rangle \right.      
\end{equation}
    
\section{Simulation Techniques}

The QWZ and BHZ models were both simulated using the Julia programming language. We used the OpenQuantumTools.jl \cite{OpenQuantumtools.jl} and QBase.jl \cite{QBase.jl} so that the available simulations, \url{https://github.com/JoshDoesAstrophysics/QWZ_BHZ_impurities}, can be read with minimal understanding of the Julia programming language. Each model and impurity type is contained in a Pluto Notebook to allow fast iteration of different site properties. Each Notebook builds a Hamiltonian and then uses the Hamiltonian for a probability time evolution.

\subsection{Lattice and Impurity Properties}

Each model version calculates a base Hamiltonian of the QWZ or BHZ model and then modifies the Hamiltonian for the impurity type being used. Each part is built and modified using defined functions that test different lattice properties. The Base Hamiltonian can be modified to have a different number of sites for the y direction through the height variable at the top of every Pluto Notebook. The Hamiltonian can also be modified to have a different number of sites for the x direction through the width variable at the top of every Pluto Notebook. The last modification to the base Hamiltonian is the potential depth for all sites, and this is left at a value of $u=-1.0$ in this paper. The base Hamiltonian is split into three summation loops correlating to the three summations in Eq. (\ref{QWZ}). Each loop is then declared a process to be executed simultaneously through multi-processing. This multi-processing step ensures that the base Hamiltonian calculation is only as slow as the slowest summation rather than the sequential operation of all three summations. The multi-processing step is the first significant speed improvement we used, and the Julia programming language allows for easy implementation.

The base Hamiltonian is modified to add in the impurity. Currently, the function that adds the impurity is unique for every model but has two main versions. Both main versions of the impurity functions can only attach impurities to the edge sites of the lattices. The first main version of the impurity function is for adding a single impurity type to the lattices. The single impurity function type has three main inputs: the edge connection, edge position $(m_{x0}, m_{y0})$, and impurity potential $(\epsilon)$. The edge connection specifies what side of the lattices the single impurity will be connected to, which can be the top, bottom, left, or right sides. The edge position specifies where the impurity is placed along an edge. The impurity potential allows for the single impurity site to have a unique potential depth, which is left at a value of $\epsilon=-0.1$ to simplify the results in this paper. The second main version of the impurity function is for adding multiple impurities to the lattices. It has five main inputs: the number of impurities, edge connections, edge positions, impurity potential, and impurity connection conjugation. The edge connection and position have the same customization as they did with the single impurity, except they are now a list for each desired impurity. The number of impurities describes how many should be added to the lattices. The impurity potential is identical to the single impurity and defines the potential depth for all impurity sites. The impurity connection conjugation is a boolean input that allows each impurity to be a complex conjugation of the standard connection type. The impurity function is included in the base Hamiltonian multi-processing step.

\subsection{Time Evolution Calculation and Plotting}

The time evolution of a wave function in the lattices is calculated using the time-dependent solution to the Schrödinger equation with a Hamiltonian given in Eq. (\ref{time_1}). The typical size of our simulated QWZ model is over 60,000 complex value elements and over 120,000 complex value elements for the BHZ model. Using the original form of Eq. (\ref{time_1}) leads to long computation times for relatively short time evolutions since each time step requires the exponential of the Hamiltonian. The simulation can handle these large exponentials since the Julia Programming language has built-in parallel processing for linear algebra operations. The code utilizes this through the \verb+BLAS.set_num_threads(#)+ command. The parallel processing step is the second significant speed improvement we used.

\begin{equation}
    \label{time_1}
    \psi(t)=e^{-i H t}\left|\psi_0\right\rangle 
\end{equation}

The simulation would still be slowed down by calculating an exponential at each time step. The time evolution is calculated in discrete time steps by simulation. Each step is then saved so that an average of all frames can be printed. The saved steps are also used to generate an animation of the time evolution. Since the time evolution is calculated discretely, we discretized Eq. (\ref{time_1}) into Eq. (\ref{time_2}). This allows for the exponential to be calculated only once for a Hamiltonian. The discretization step is the third significant speed improvement we used and is the most effective. The discretization step significantly reduces the overhead of each time step so that a non-discretized 2000-step simulation could be scaled up to a discretized 2 million time steps for the same computation time. With this optimization to the simulation, computer memory size becomes the main computational blockade for any large-scale time evolution. Note that for Eq. (\ref{time_2}), a $\Delta t\neq1$ produces slightly different results over the time evolution, The further $\Delta t$ is from $1$, the probabilities for a time evolution deviate more as well. Only $\Delta t=1$ simulations are shown in the results section since the probabilities closely match simulations using Eq. (\ref{time_1}).

\begin{equation}
    \label{time_2}
    \psi(t+\Delta t)=e^{-i H \Delta t}|\psi(t)\rangle
\end{equation}

\section{\label{sec:level1}Simulation Results}

\begin{figure}
    \centering
        \includegraphics[width=0.45\textwidth]{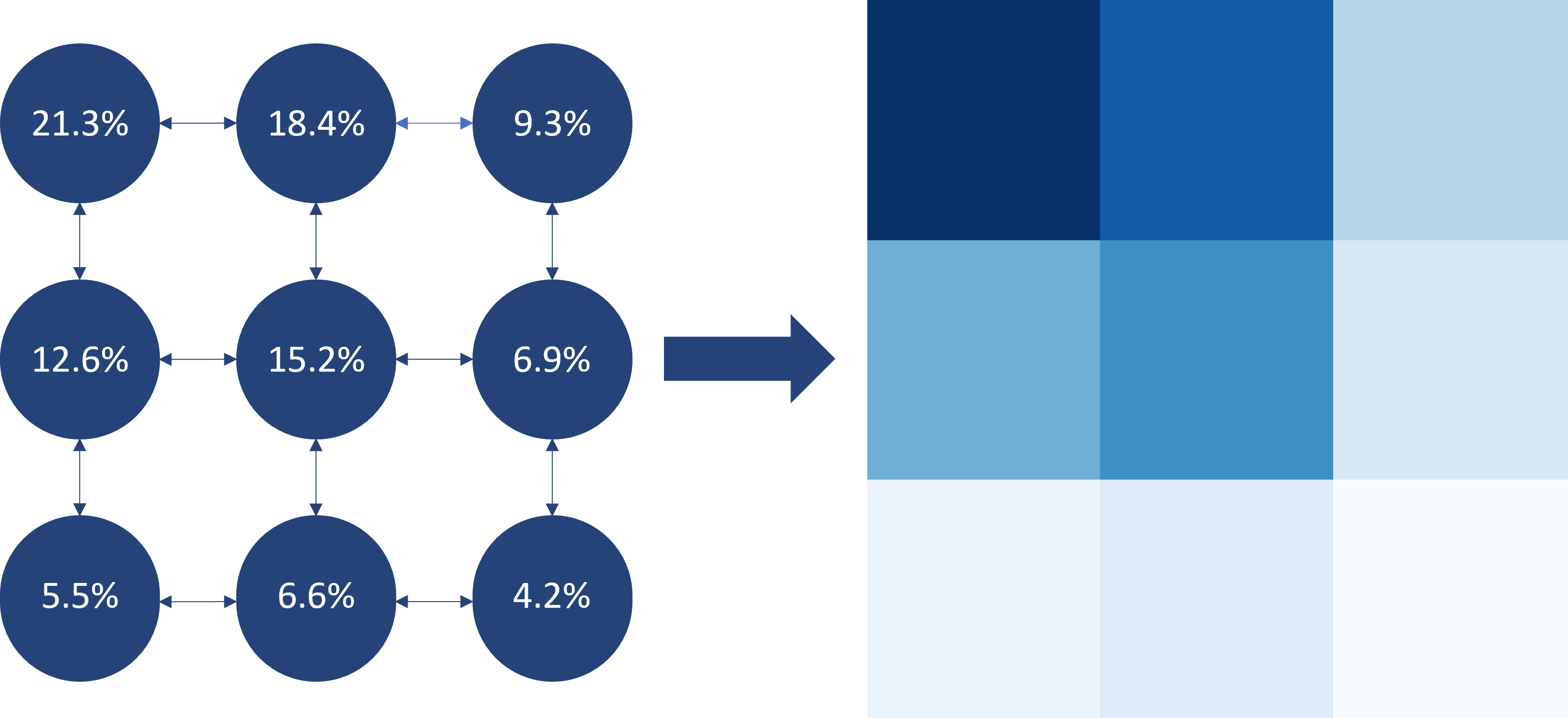}
    \caption{\label{fig:heat}The left image shows the probability for each site for a three-by-three (3x3) QWZ lattice with no impurity where the paramagnetic state probabilities for each site are combined into one value. The right block shows how these probabilities are positionally mapped onto a heatmap where higher probabilities are darker shades of blue.}
\end{figure}

Our first attempt at simulating the finite QWZ and modified BHZ model demonstrates results consistent with the infinite case of each model \cite{Asboth16}. The results are analyzed through an animation of the time evolution and an average of the time evolution as the output of the simulation. Both outputs are a heatmap representation of the wave function probabilities in the lattice. The modified BHZ output shows two heatmaps, each for the time-forward pair and its time-reversal pair. Figure \ref{fig:heat} shows how the probability is mapped from a three-by-three QWZ model with no impurity where the wave function has begun to propagate. For the simplicity of discussing the results from the simulations, all simulations will have a $u = -1.0$, $\epsilon = -0.1$, and $\sigma_w = \sigma_x$ defined in Equations \ref{QWZ} and \ref{impurity}.

\begin{figure}[b]
    \centering
        \includegraphics[width=0.5\textwidth]{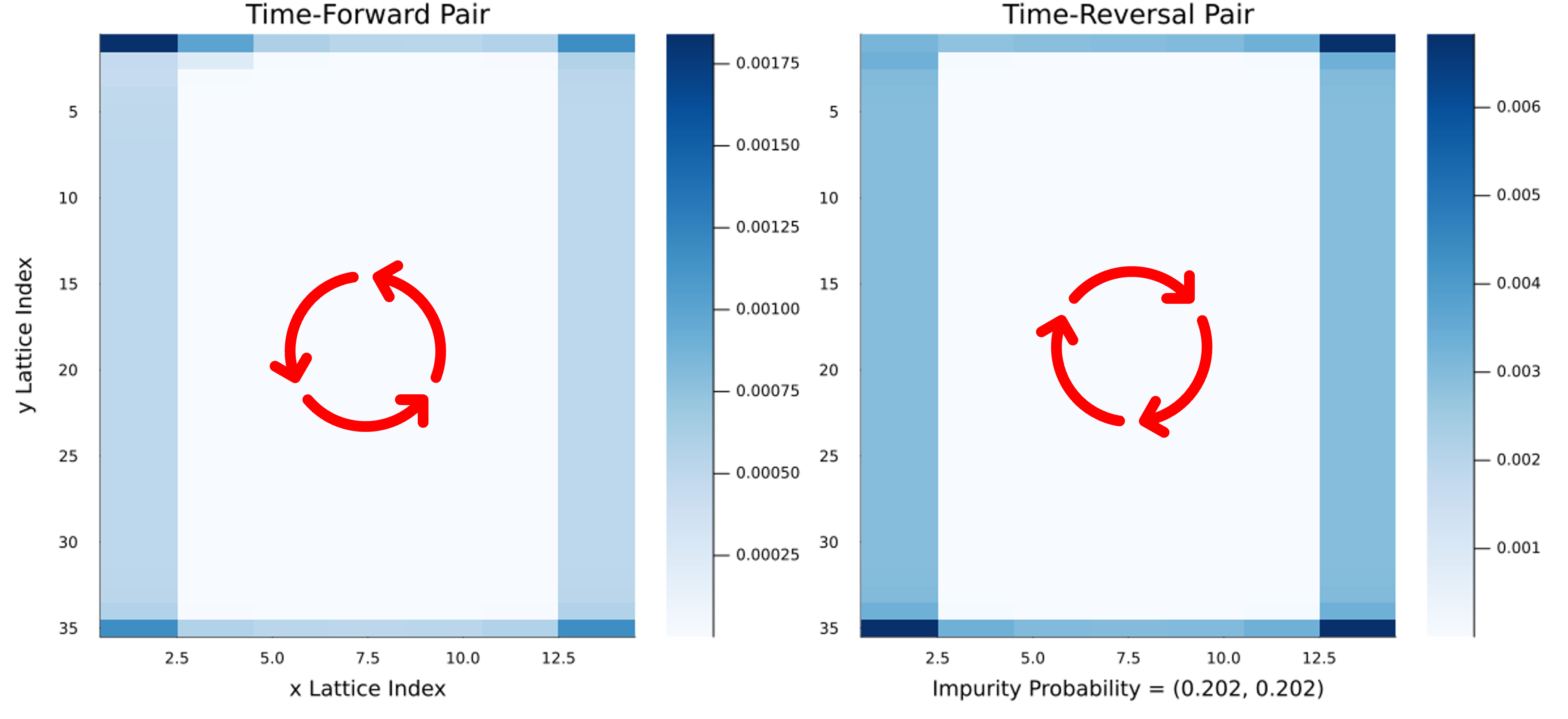}
    \caption{\label{fig:corner}The figure depicts a 7x35 modified BHZ lattice where an impurity is connected horizontally, see $\sigma_w = \sigma_x$ for Eq. (\ref{impurity}), to the top left corner. Both pairs show that simulating a million time steps the probability of the wave function is along the edge of both lattice pairs, and avoids the bulk of central sites.}
\end{figure}

\subsection{Impurity Corner Lattice Scattering}

When the impurity is connected to a corner site of a lattice, corner lattice scattering behavior is observed. The infinite case of the models has an eigenstate associated with directional scattering along the edges. This state is observed in our simulation of finite versions of the model. When an impurity is attached to a corner and contains the initial probability, the time simulation shows that wave scattering moves counterclockwise along the edge of the time-forward pair and clockwise along the edge of the time-reversal pair. Figure \ref{fig:corner} shows how the probability evolves over the average of the simulation over a long period of time for corner lattice scattering behavior. Corner lattice scattering behavior has the unique property where scattering is observed to immediately occur in both the time-forward pair and the time-reversal pair from any of the eight corners without symmetry breaking.

\begin{figure}[b]
    \centering
        \includegraphics[width=0.5\textwidth]{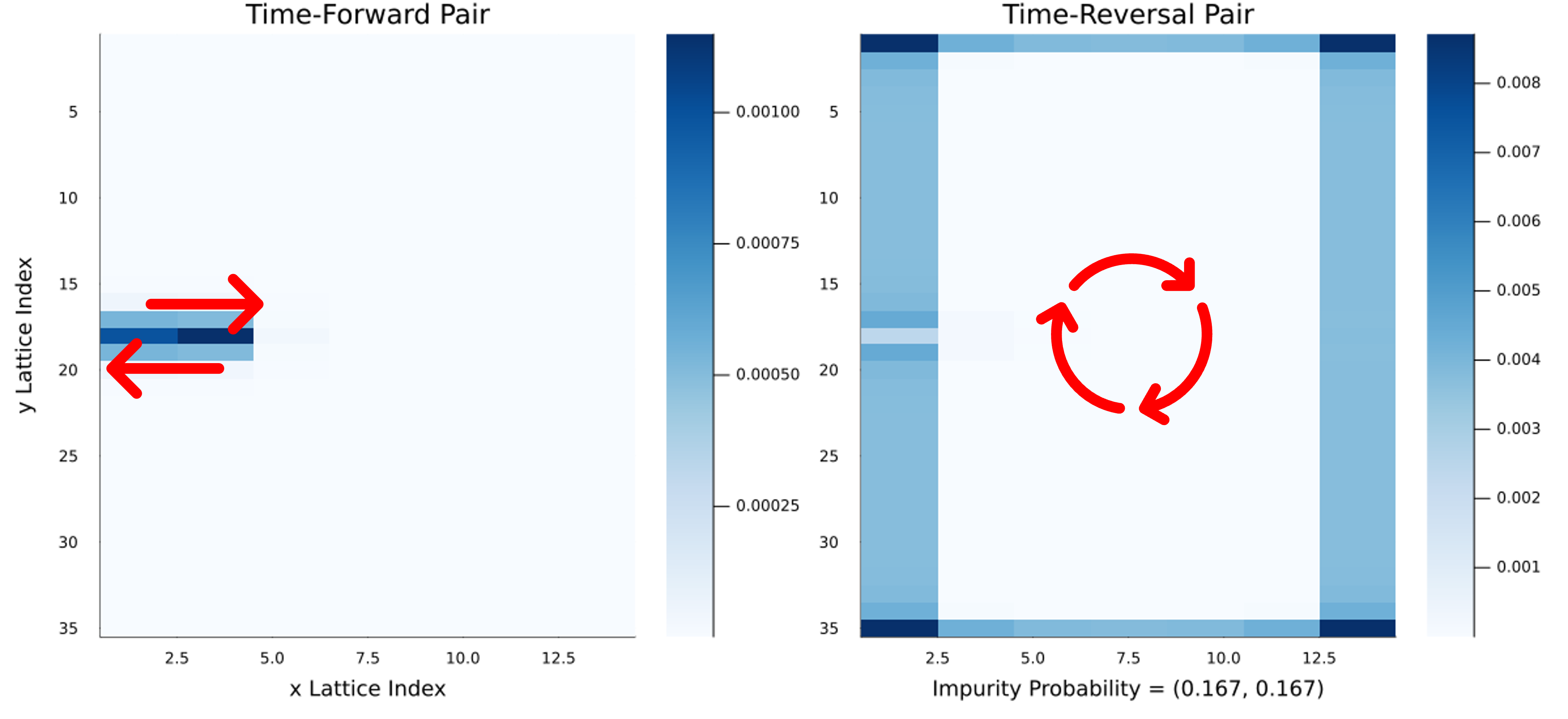}
    \caption{\label{fig:edgeLeft}The figure depicts a 7x35 modified BHZ lattice where an impurity is connected horizontally to the middle left edge. Both pairs show that simulating a million time steps the probability of the wave function is stuck near the impurity for the time-forward pair and scattering for the time-reversal pair.}
\end{figure}

\subsection{Impurity Edge Lattice Scattering and Absorption}

\begin{figure}
    \centering
        \includegraphics[width=0.5\textwidth]{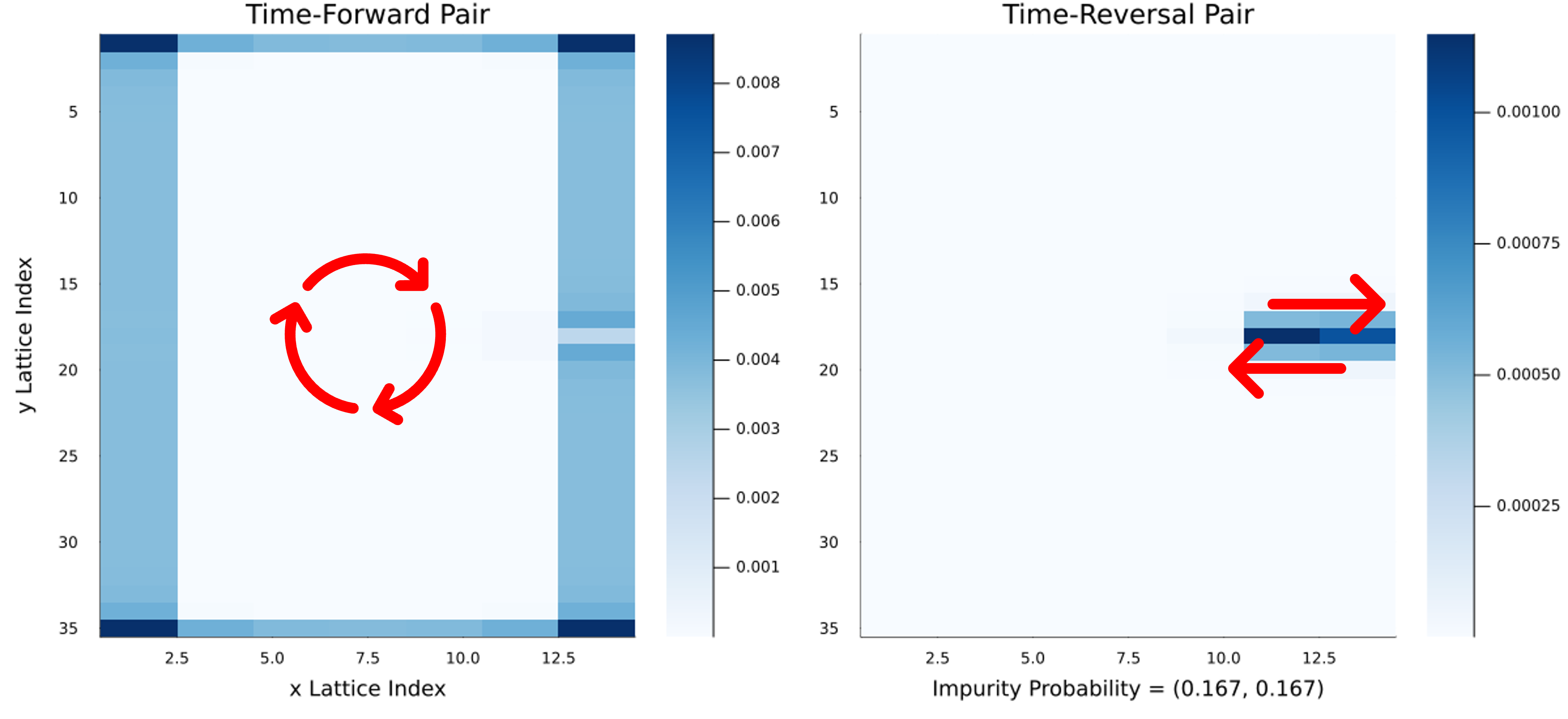}
    \caption{\label{fig:edgeRight}The figure depicts a 7x35 modified BHZ lattice where an impurity is connected horizontally to the middle right edge. Both pairs show that simulating a million time steps the probability of the wave function is scattering for the time-forward pair and absorbing near the impurity for the time-reversal pair.}
\end{figure}

Edge lattice scattering and absorption behavior are observed when the impurity is connected to a center edge of each lattice. Each side and lattice-type connection produces a different result for the probability function. Figure \ref{fig:edgeLeft} shows how the probability evolves over the average of the simulation over a long period of time for edge lattice scattering and absorption behavior. Figure \ref{fig:edgeLeft} has the connection on the left side of each lattice, and the scattering matches the behavior of the corner lattice scattering for the time-forward, but the time-reversal pair has absorption behavior. Figure \ref{fig:edgeRight} where the impurity is connected to the right of the lattices, the time-reversal pair has scattering behavior, and the time-forward pair has absorption behavior. 

\begin{figure}[b]
    \centering
        \includegraphics[width=0.5\textwidth]{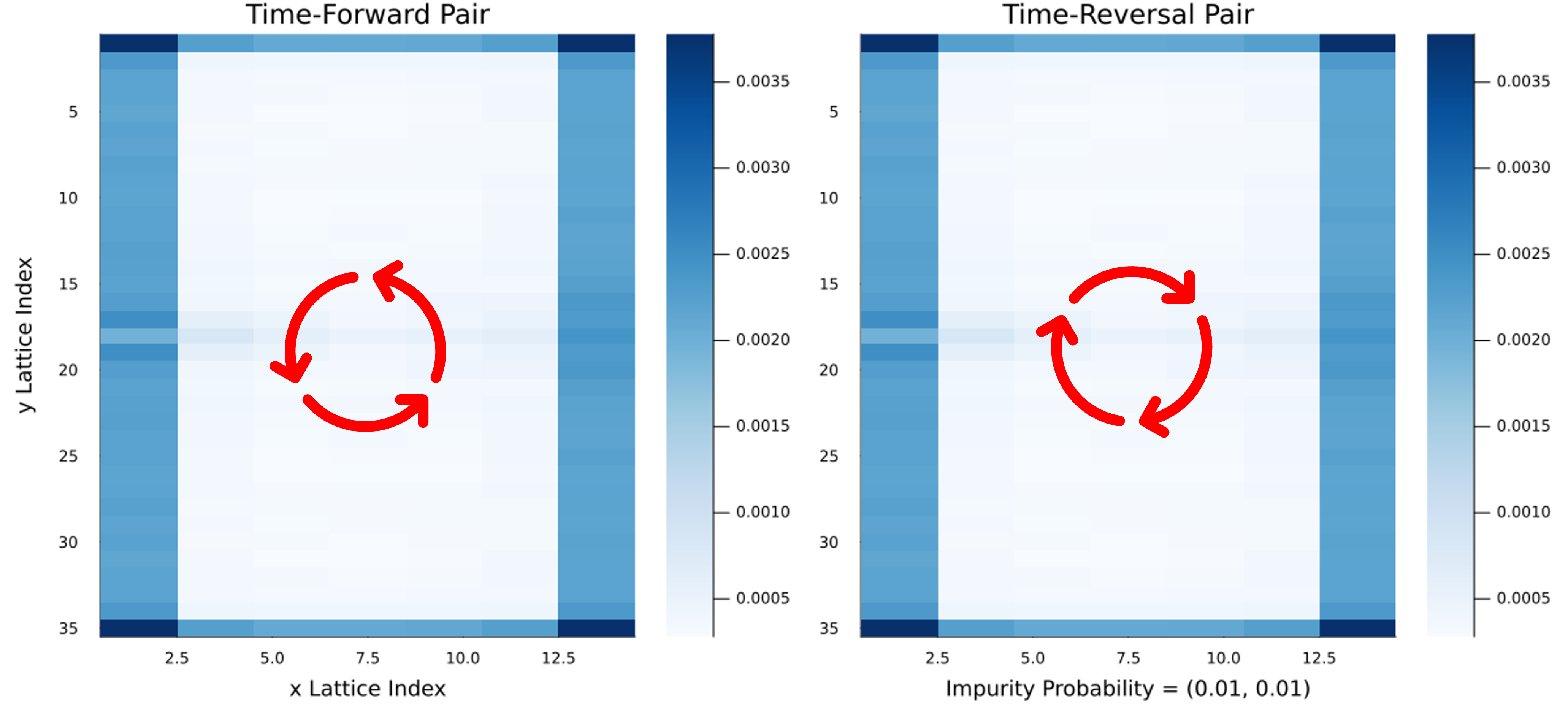}
    \caption{\label{fig:anti-symmetric}The figure depicts a 7x35 modified BHZ lattice where an impurity is connected horizontally to the middle left edge and the time forward pair has the complex conjugate of its impurity connection. Both pairs show that simulating a million time steps the probability of the wave function is scattering for both time pairs.}
\end{figure}

\subsection{Impurity Anti-Symmetry Scattering}

Anti-symmetry lattice scattering behavior was only observed when the impurity was connected to a center edge of each lattice, and the connection for the impurity was the complex conjugate of a normal connection used for the previous two sections, with the following Hamiltonian in Eq. (\ref{anti}).
\begin{equation}    
    \label{anti}
    \widehat{H}_{\rm BHZ} = \left( \widehat{H}_{\rm QWZ} +\widehat{H}_{\rm IMP}\right)\otimes \left|\uparrow\rangle \right. +  \left( \widehat{H}_{\rm QWZ} +\widehat{H}_{\rm IMP}^*\right)^* \otimes \left|\downarrow\rangle \right.      
\end{equation}

Figure \ref{fig:anti-symmetric} shows how the type of connection controls whether or not an edge lattice connection has corner lattice scattering behavior, where there is simultaneous scattering behavior for both lattices. This demonstrates that scattering in the model is controlled by the configuration of the imaginary component in the x-direction and the location of the impurity.

\section{Discussions}

Our simulations of a two-dimensional paramagnetic semiconductor should be treated as a building point for simulating more physically realistic lattices. The three scattering and absorption behaviors observed demonstrate the exciting mechanics that can be observed with different modifications to the lattices using impurities. The results we display here are only a tiny section of the possible behaviors present for this lattice. The code for simulation presents the ability to connect the impurity to any edge position of the lattices for testing different results and adding multiple impurities to the modified BHZ model. Meanwhile, the results presented only look at connections on the corners, left central edge, and right central edge. Future investigations with these simulations can expect to find unique scattering and absorption behaviors. These behaviors may be useful for controlling the current through the edge of the lattice using magnetic impurities \cite{Liu09, Winters22}. The unique scattering and absorption behaviors can be used in the input of the design of useful electronic systems and components. 

\begin{acknowledgments}

We would like to recognize the generosity of the Seitz Family and Butler University's Liberal Arts and Sciences for funding the research in this paper. Acknowledgments also extended to the Quantum Dynamics Group at Osaka Metropolitan University for their invitation to work at their university while conducting this research. Dr. Savannah Garmon of Osaka Metropolitan University and Dr. Dan Kosik of Butler University also provided their insight for this paper. Former Butler student Dejuan Winters is recognized for beginning the work on these models while at Butler University.  

\end{acknowledgments}

\bibliography{apssamp}

\end{document}